\documentclass[a4paper,10pt]{article}
\usepackage{jheppub}
\usepackage[utf8]{inputenc}
\usepackage{amsmath}
\usepackage{amssymb}
\usepackage{tikz-feynman}%[compat=1.1.0]
\usepackage{xcolor}

\title{\boldmath To the sphere and back again: de Sitter infrared correlators at NTLO in 1/N }

\author[a]{Diana L\'opez Nacir}
\author[b]{Francisco D. Mazzitelli}
\author[c]{Leonardo G. Trombetta}

\affiliation[a]{Departamento de F\'\i sica and IFIBA, FCEyN UBA, Facultad de Ciencias Exactas y Naturales, 
 Ciudad Universitaria, Pabell\' on I, 1428 Buenos Aires, Argentina}
\affiliation[b]{Centro At\'omico Bariloche, Instituto Balseiro and CONICET,\\ Comisi\'on Nacional de Energ\'\i a At\'omica, Av. Bustillo 9500, R8402AGP Bariloche, Argentina.}
\affiliation[c]{Scuola Normale Superiore, Piazza dei Cavalieri 7, 56126, Pisa, Italy\\
INFN – Sezione di Pisa, 56200, Pisa, Italy}

\emailAdd{dnacir@df.uba.ar}
\emailAdd{fdmazzi@cab.cnea.gov.ar}
\emailAdd{leonardo.trombetta@sns.it}

\abstract{   We analyze the infrared behavior of the two and four-point functions for the massless $O(N)$ model in Lorentzian de Sitter  spacetime, using the $1/N$ expansion. 
Our approach is based in the study of the Schwinger-Dyson equations on the sphere (Euclidean de Sitter space), using the fact that the infrared behavior in Lorentzian spacetime is determined by the pole structure of the Euclidean correlation functions. We compute the two-point function up to the NTLO in $1/N$, and show that in the infrared it behaves as the superposition of two massive free propagators with effective masses of the same order, but not equal to, the dynamical mass $m_{dyn}$. We compare our results with
those obtained using other approaches, and find that they are equivalent but retrieved in a considerably simpler way. 
We also discuss the infrared behavior of the equal-times four-point functions. }

\begin{document}

\maketitle

\section{Introduction}

The analysis of the infrared (IR) behavior of correlation functions for interacting fields in de Sitter (dS) spacetime is of high interest in the context of semiclassical and quantum gravity. It has been shown that the loop expansion breaks down at large dS-invariant distances for light fields, due to seculary growing corrections. It is particularly compelling for massless  fields, where the free two-point function shows a non-dS invariant behavior, and does not vanish for largely separated points. 

Several non-perturbative techniques have been
developed to cure this problem:  the stochastic approach \cite{stoch}, Hartree approximation \cite{Hartree}, $1/N$ expansion \cite{1N}, renormalization group equation
\cite{SerreauRG}, exact treatment of the zero-mode in Euclidean space \cite{HuOconnor, Rajaraman, BenekeMoch}, including partial resummations of the non-zero modes 
\cite{nos}, etc. For a recent review on the IR behavior of quantum fields in inflationary cosmology see Ref.\cite{BeiLok}. 

In this paper, we will be concerned with
the analysis of the problem in the context of the $O(N)$ model, in the large $N$ limit. It is well known that, to leading order, the propagators of the interacting fields 
are given by free massive propagators, with a non-perturbative dynamical mass $m_{dyn}^2=\sqrt{\lambda/(2 V_d)}$, where $\lambda$ is the coupling constant and
$V_d$ the volume of the $d$-sphere. As a free massive propagator decays exponentially in the IR, 
the non-perturbative result restores the usual dS-invariant behavior of  a massive two-point function. However, there are still secular contributions in each individual diagram that need to be resummed in a consistent way \footnote{See for instance \cite{Akhmedov:2019cfd} for a discussion of different types of secular effects.}.

Our main goal here is to compute the NTLO $1/N$ corrections to the correlation functions. Although this problem has been studied before \cite{Serreau2,nos}, we will provide an alternative, and technically simpler
approach that may be used to generalize the results to compute NNTLO corrections.  
The main idea is the following:    As shown in \cite{HM&M}  (see also \cite{Korai:2012fi}) the correlators of an  interacting massive theory, computed using in-in perturbation theory in the
expanding cosmological patch of dS space, and for which the free propagators are taken to be those of the Bunch-Davies vacuum, coincide with those obtained by analytic continuation from Euclidean dS; i.e., with the correlators in the fully interacting theory on the sphere. Since at  leading order in $1/N$, the two-point correlation functions of the theory correspond to massive propagators with the  self-consistent dynamical mass, the next to leading order corrections in $1/N$ are computed using these massive two-point functions in the Feynman diagrams. We can therefore apply the previous results obtained for massive theories. In this approach, dS invariance is maintained, and therefore one can go from the sphere to Lorentzian dS spacetime by analytic continuation of the isometry invariant. 
 The propagators on the sphere admit an expansion in spherical harmonics.
  When continued back to dS, the IR\footnote{With IR we mean the leading behaviour, after analytical continuation, at large values of the isometry invariant.}  behavior is determined by the presence of poles in the complex $L$-plane ($L$ being the angular momentum) \cite{M&M, Hollands}. Therefore, to compute the NTLO, we will solve the 
Schwinger-Dyson equation (SDE) on the sphere, using as  input the above mentioned relation between poles and IR behavior. Then we will rotate back the propagators to dS to find the NTLO behavior.

\section{SDE on the sphere}

For the $O(N)$ model, with Euclidean action
\begin{equation}
 S = \int d^d x \sqrt{g} \left[ \frac{1}{2} \phi_a \left( -\square + m_{dS}^2 \right) \phi_a + \frac{\lambda}{8N} ( \phi_a \phi_a )^2 \right], \label{Eaction}
\end{equation}
the SDE in Euclidean signature reads \cite{Serreau2}
\begin{equation}
 \left( -\square + m_{dS}^2 \right) G(x,x') + \int_z \Sigma(x,z) G(z,x') = \delta(x,x'),
\end{equation}
where $G(x,x')$ is the full propagator of the theory. Here, all the $O(N)$-index structure has already been simplified assuming $G_{ab} = \delta_{ab} G$ (i.e. symmetric phase), and boils down to the specific factors of $N$ found in the coming expressions. The self-energy at NTLO in $1/N$ is given by
\begin{equation}
 \Sigma(x,x') = \frac{\lambda (N+2)}{2N} [G] \delta(x,x') + \frac{\lambda}{N} G(x,x') I(x,x').
\end{equation}
In this expression, $I(x,x')$ takes care of resumming all the diagrams that contribute at NTLO in $1/N$. Some such diagrams are shown in Fig.~\ref{fig:1}. Indeed, each time an extra bubble is added there is a factor $N$ from the trace over the loop, while there is also a $1/N$ factor coming from the vertex, giving an overall contribution of the same order regardless the number of bubbles in the chain. Therefore, we say that $I(x,x')$ corresponds to a resummed bubble-chain, which satisfies the following self-consistent equation
\begin{equation} \label{Ieuc}
I(x,x') = \Pi(x,x') + \int_z \Pi(x,z) I(z,x'),
\end{equation}
where $\Pi(x,x')$ is the single bubble,
\begin{equation}\label{single-bubble}
 \Pi(x,x') = - \frac{\lambda}{2} G(x,x')^2.
\end{equation}
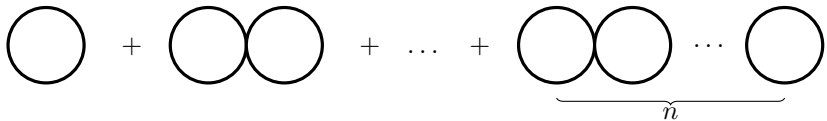
\begin{figure}[h!]
 \centering
\begin{tikzpicture}[anchor=base, baseline=-0.1cm]
  \begin{feynman}
    \vertex (a);        
    \vertex at ($(a) + (0cm, 0cm)$) (n) [blob,very thick,fill=white,minimum size=1cm] {};
  \end{feynman}
\end{tikzpicture}
\quad+\quad
\begin{tikzpicture}[anchor=base, baseline=-0.1cm]
  \begin{feynman}
    \vertex (a);
    \vertex at ($(a) + (0cm, 0cm)$) (n1) [blob,very thick,fill=white,minimum size=1cm] {};
    \vertex [right=1cm of n1] (n2) [blob,very thick,fill=white,minimum size=1cm] {};
  \end{feynman}
\end{tikzpicture}
\quad+\quad\dots\quad+\quad
\begin{tikzpicture}[anchor=base, baseline=-0.1cm]
  \begin{feynman}
    \vertex (a);
    \vertex at ($(a) + (0cm, 0cm)$) (n1) [blob,very thick,fill=white,minimum size=1cm] {};
    \vertex [right=1cm of n1] (n2) [blob,very thick,fill=white,minimum size=1cm] {};
    \vertex [right=1cm of n2] (n3) {\(\dots\)};
    \vertex [right=1cm of n3] (n4) [blob,very thick,fill=white,minimum size=1cm] {};
    
    \draw [decoration={brace}, decorate] ($(n4) + (0cm, -0.7cm)$) -- ($(n1) + (0cm, -0.7cm)$) node[pos=0.5, below] {  \(n\)};    
  \end{feynman}
\end{tikzpicture}
\caption{Diagrams contributing to $I(x,x')$ at the same order in $1/N$.}
\label{fig:1}
\end{figure}

The analytical continuation of dS spacetime to Euclidean signature has the metric of a $d$-sphere of radius $H^{-1}$
\begin{equation}
 ds^2 = H^{-2} \left[ d \theta^2 + \sin(\theta)^2 d\Omega^2 \right], \label{d-sphere}
\end{equation}
with $\theta = H\tau$. Exploiting its symmetry and compactness, any function of two points $F(x,x')$ can be expanded in the d-dimensional spherical harmonics, 
\begin{equation}
 F(x,x') = V_d \sum_{\vec{L}} F_L \, Y_{\vec{L}}(x) Y^*_{\vec{L}}(x'),
\end{equation}
such that the $F_L$ only depend on $L \equiv |\vec{L}|$. The $V_d$ factor is conventional. Transforming the previous equations swaps the convolutions for simple products (and viceversa).  The ``momentum''-space SDE, now algebraic, reads:
\begin{equation}
 \left[ L(L+d-1) + m_{dS}^2 \right] G_L + V_d \, \Sigma_L G_L = \frac{1}{V_d},
\end{equation}
where
\begin{eqnarray} 
 \Sigma_L &=& \frac{\lambda (N+2)}{2N V_d} [G] + \frac{\lambda}{N} \bar{\Sigma}_L, \\
I_L &=& \Pi_L + V_d \, \Pi_L I_L,  \label{I_L-eq} \\
 \Pi_L &=& - \frac{\lambda}{2} \rho_L.
\end{eqnarray}
Here we have defined $\bar{\Sigma}(x,x') \equiv G(x,x') I(x,x')$ and $\rho(x,x') \equiv G(x,x')^2$. These equations are algebraic and easily solved once all objects are known. The tricky part is computing the coefficients of the spherical harmonic expansions of $\Sigma_L$ and $\rho_L$. 

Notice that the first term of the self-energy is local and thus can be absorbed in the mass, 
\begin{eqnarray}
 M^2 &=& m_{dS}^2 + \frac{\lambda(N+2)}{2N}[G] \notag \\
 &=& m_{dS}^2 + m_0^4 \left(1+\frac{2}{N} \right) V_d [G],
\end{eqnarray}
where in the second line we are expressing the coupling $\lambda$ in terms of the quantity $m_0^2 = \sqrt{\lambda/2V_d}$, in anticipation to the LO result. All further instances of the coupling will be expressed in this manner.  Therefore the equations read
\begin{equation}
 \left[ \chi_L + M^2 \right] G_L + \frac{2 m_0^4 V_d^2}{N} \, \bar{\Sigma}_L G_L = \frac{1}{V_d}, \label{SD-eqs-EdS}
\end{equation}
where it is explicit that the bubble-chain contributions (contained in $\bar{\Sigma}$) are of NTLO in $1/N$ and we also defined the shorthand  $\chi_L \equiv H^2 L(L+d-1)$. Since $\bar{\Sigma}_L$ depends on $G_L$, this is a self-consistent equation for $G_L$.  However, since we are looking for the NTLO corrections, we need only to know the part of $\bar{\Sigma}_L$ that is LO in $1/N$, and therefore for its computation we can use the LO part of $G_L$ as well, which is well known to be a free propagator with a dynamical mass $m_{dyn}$ (which has yet to be determined), i.e. $G_L \simeq \Delta_L^{(m_{dyn})}$.  

\section{Solving the SDE in the IR}

\subsection{Poles and IR behaviour}

A free propagator of mass $m$ in dS has the following asymptotic behaviour in the IR,
\begin{equation}\label{IRfree}
 \Delta^{(m)}(x,x') = \frac{r^{-m^2/(d-1)H^2}}{V_d m^2} \left[ 1 + \frac{m^2}{2r (d-1) H^2} + \dots \right],
\end{equation}
where $r(x,x')$ is the dS invariant distance, which in Lorentzian signature it is free to grow boundlessly as $x$ and $x'$ are separated. The corresponding Euclidean counterpart on the sphere in ``momentum''-space has the following representation
\begin{equation}
\Delta_L = \frac{1}{V_d \left( \chi_L + m^2 \right)}.
\end{equation}
The two can be shown to be related by analytical continuation, even for arbitrary distances, through their exact expressions (see for example Ref. \cite{Garbrecht:2013coa}). However, for general functions we must rely on the connection between the leading IR behavior of any given two-point function $F(x,x')$ in dS, and the pole structure in the region $Re(L) \leq 0$ of the momentum-space transform $F_L$ of the corresponding Euclidean function. Indeed, as shown in Refs.~\cite{M&M, Hollands}, this behavior is $r^{-|Re(\bar L)|}$ with $\bar L$ the position of the pole  which lies closer to $L=0$. The particular example of the free propagator above is illuminating:  $\Delta_L$ has   poles at $L = - z_{\pm} = - \frac{d-1}{2} \pm \nu$, with $\nu = \sqrt{\frac{(d-1)^2}{4} - \frac{m^2}{H^2}}$. For $m^2 \ll H^2$, these are
\begin{eqnarray}
 z_+ &\simeq& \frac{m^2}{(d-1) H^2} \ll 1, \\
 z_- &\simeq& d-1 - \frac{m^2}{(d-1)H^2 } \sim \mathcal{O}(1).
\end{eqnarray}
The IR behaviour is then dominated by the residue at $L = - z_{+}$, as shown by Eq.\eqref{IRfree} (note that for massless fields there is a pole at $L=0$, that leads to a non-decaying behavior at large separations that goes as $\log(r)$).

We will use this property to study the IR behaviour in dS from the analytical continuation of a two-point function $F(x,x')$ on the sphere,  finding the poles of the corresponding $F_L$ that lie in the region $-1 \ll Re(L) \leq 0$ in the complex $L$-plane. As long as there is no pole at $L=0$, we expect a decay at large separations of the points $x$ and $x'$.
Our goal is therefore to study the pole structure of
$G_L$ in the aforementioned region of the angular momentum complex plane.

%Here we are dropping terms that decay at least with an extra $1/r$ factor, which would correspond to a ``mass'' of order $H$. 

\subsection{An approximation for the self-energy}

We will now estimate 
both $\rho_L$ and $\bar{\Sigma}_L$ in the IR. We remind that the coefficients $\rho_L$ are those of  the square of the propagator. 
Consider the product of two free propagators in dS with masses $m_1$ and $m_2$, in the same pair of spacetime points $x$ and $x'$.  The IR asymptotic behaviour of this product is simply
\begin{eqnarray} \label{IR-DeltaDelta}
 \Delta^{(m_1)}(x,x') \Delta^{(m_2)}(x,x') &=& \frac{r^{-(m_1^2 + m_2^2)/(d-1)H^2}}{V_d^2 m_1^2 m_2^2} \left[ 1 + \frac{(m_1^2 + m_2^2)}{2r (d-1) H^2} + \dots \right] \nonumber \\
 &\simeq& \frac{(m_1^2 + m_2^2)}{V_d m_1^2 m_ 2^2} \Delta^{(\sqrt{m_1^2 + m_2^2})}(x,x') ,
\end{eqnarray}
where in the second line we have conveniently expressed it in terms of a single propagator with a squared-mass equal to the sum of the individual squared-masses, with an appropriate coefficient. This translates easily to a corresponding representation on the sphere, 
\begin{equation}
\left[\Delta^{(m_1)} \Delta^{(m_2)}\right]_L = \frac{(m_1^2 + m_2^2)}{V_d^2 m_1^2 m_ 2^2} \, \frac{1}{\left( \chi_L + m_1^2 + m_2^2 \right)}.
\end{equation}
What we are saying here is that the fact that two dS expressions are similar in the IR (i.e. first and second lines of Eq.~\eqref{IR-DeltaDelta}) is equivalent to their Euclidean counterparts having similar pole structures in $L$ in the region $-1 \ll Re(L) \leq 0$. This is the sense in which we do an IR approximation of an Euclidean expression that we intend to analytically continue to dS. Recalling that the propagators in the SDE are the exact propagators $G_L$, which at LO in $1/N$ are just free propagators with a dynamical mass $m_{dyn}$, allows us to use this result to obtain
\begin{equation}
 \rho_L \simeq \frac{2}{V_d m_{dyn}^2} \Delta_L^{(\sqrt{2} m_{dyn})}. \label{rho_L}
\end{equation}
This in turn allows us to find the resummed bubble-chain by simply algebraically solving Eq.~\eqref{I_L-eq}, giving
\begin{equation}
 I_L = \frac{\Pi_L}{1 - V_d \Pi_L} \simeq - \frac{2 m_0^4}{m_{dyn}^2} \frac{\Delta_L^{(\sqrt{2} m_{dyn})}}{\left[1 + \frac{2V_d m_0^4}{m_{dyn}^2} \Delta_L^{(\sqrt{2} m_{dyn})}\right]} = - \frac{2m_0^4}{m_{dyn}^2} \Delta_L^{(\sqrt{2} \bar{m}_{dyn})}  , 
\end{equation}
where we have defined the shorthand $\bar{m}_{dyn}^2 = m_{dyn}^2 + \frac{m_0^4}{m_{dyn}^2}$. This result can be analytically continued to dS by just exploiting the fact it looks like a free massive propagator, 
\begin{equation} \label{IeucRes}
I(x,x') \simeq - \frac{2m_0^4}{m_{dyn}^2} \Delta^{(\sqrt{2} \bar{m}_{dyn})}(x,x'),
\end{equation}
which is already an interesting result. 

Moving on with the SDE, in order to estimate $\bar{\Sigma}^{(LO)}(x,x') \simeq \Delta^{(m_{dyn})}(x,x') I(x,x')$ in the IR we now perform again a step in dS by exploiting the fact that its expression is given by the product of two free massive propagators,
\begin{eqnarray}
 \bar{\Sigma}^{(LO)}(x,x') &\simeq& - \frac{2m_0^4}{m_{dyn}^2} \Delta^{(m_{dyn})}(x,x') \Delta^{(\sqrt{2} \bar{m}_{dyn})}(x,x') \nonumber \\
 &\simeq& - \frac{m_0^4 (m_{dyn}^2 + 2 \bar{m}_{dyn}^2)}{V_d m_{dyn}^4 \bar{m}_{dyn}^2} \Delta^{(\sqrt{m_{dyn}^2 + 2 \bar{m}_{dyn}^2})}(x,x'), \label{Self-E-IR-1}
\end{eqnarray}
which then can easily be taken back to the sphere,
\begin{eqnarray}
 \bar{\Sigma}^{(LO)}_L \simeq - \frac{m_0^4 (m_{dyn}^2 + 2 \bar{m}_{dyn}^2)}{V_d m_{dyn}^4 \bar{m}_{dyn}^2} \Delta_L^{(\sqrt{m_{dyn}^2 + 2 \bar{m}_{dyn}^2})}
 \simeq -\frac{5}{2V_d^2} \frac{1}{\left( \chi_L + 5m_0^2 \right)} .\label{SigmaL}
 %-\frac{5m_0^4}{\lambda}. 
\end{eqnarray}
This is the main ingredient needed for solving the SDE on the sphere. Note that, since $ \bar{\Sigma}_L$ is multiplied by $1/N$ in Eq.\eqref{SD-eqs-EdS} as discussed before, in the last step of the above equation we made the replacements $m^2_{dyn} \simeq \bar  m^2_{dyn}/2 \simeq m_0^2$. Note that, when analytically continued, the self-energy and the resummed bubble-chain decay as free propagators with squared masses $5m_0^2$ and $4m_0^2$ respectively.

\subsection{The propagator up to NTLO in $1/N$}

With these ingredients we can   finally solve the SDE for $G_L$ at NTLO in $1/N$. From Eqs.~\eqref{SD-eqs-EdS} and \eqref{SigmaL} we obtain
\begin{equation}
 G_L = \frac{1}{V_d \left[\chi_L + M^2 - \frac{5 m_0^4}{N}\frac{1}{(\chi_L + 5 m_{0}^2)} \right]} =  \frac{\chi_L+5m_0^2}{V_d \left[(\chi_L + M^2)(\chi_L+5m_0^2) - \frac{5 m_0^4}{N} \right]} , \label{SD-explicit}
\end{equation}
where we used the explicit expression for $\Delta_L$. We can rewrite this expression using a partial fraction decomposition
\begin{equation}
 G_L = \frac{c_+}{V_d(\chi_L + m_+^2)} + \frac{c_-}{V_d(\chi_L + m_-^2)}, \label{SD-2props}
\end{equation}
with a proper choice of coefficients $c_+$ and $c_-$, and masses $m_+^2$ and $m_{-}^2$, which must satisfy
\begin{subequations}\label{c-coeffs}
\begin{eqnarray}
 c_+ + c_- &=& 1,\\
 c_+ m_-^2 + c_- m_+^2 &=& 5 m_0^2.
\end{eqnarray} 
\end{subequations}
Here the masses are the roots of the denominator in Eq.\eqref{SD-explicit} with their signs reversed.

From Eq.\eqref{SD-2props} we already see that the corrected propagator can be approximated by a linear combination of two free propagators, with masses $m_\pm$. To determine these masses, as well as the coefficients $c_\pm$, $M^2$ must be (self-consistently) computed.  We note that
\begin{equation}
M^2= m_{dS}^2 + m_0^4 \left(1+\frac{2}{N} \right) V_d (G_0 + [\hat{G}])\, ,
\end{equation}
where $[\hat{G}]$ contains all contributions with $L\neq 0$, and it is a divergent quantity. The usual mass renormalization leads to a finite expression
% allows  divergence can be absorbed into the bare mass $m^2_{dS}$ and therefore
\begin{equation}
M^2=m_{dS}^{2}\vert_{ren} + m_0^4 \left(1+\frac{2}{N} \right) V_d (G_0 + [\hat{G}]_{ren} )\, .
\end{equation}
In what follows we will work in the massless case $m_{dS}^{ 2}\vert_{ren}=0$. Moreover, notice that as long as $m_{dyn}^2 \ll H^2$, which is our working assumption, we can also neglect $[\hat{G}]_{ren} \sim H^2$ with respect to $G_0 = 1/V_dm_{dyn}^2 \gg H^2$. Since all the masses in the problem (when $m_{dS}^{ 2}\vert_{ren}=0$) end up being proportional to $m_0^2 \sim \sqrt{\lambda} H^2$, this statement can be made parametrically accurate by demanding $\lambda \ll 1$.
 
Let us focus on the $L=0$ component of the SDE~\eqref{SD-explicit}, which is a self-consistent equation for $G_0$ (or for the dynamical mass $m^2_{dyn}$). We have
\begin{equation}
 G_0 \equiv \frac{1}{V_d m_{dyn}^2} =\frac{1}{V_d}\frac{1}{\left[m_0^4 (1+\frac{2}{N})V_d G_0-\frac{m_0^2}{N}\right]}, 
 \end{equation}
whose solution up to NTLO in $1/N$ is
\begin{equation}
 m_{dyn}^2 = m_0^2 \left( 1 + \frac{1}{2N} \right), \label{mdyn}
\end{equation}
a well known result. 

Knowing the value of $m_{dyn}^2$, we can determine $M^2$. As already mentioned, the masses $m^2_\pm$ are minus the roots of the second degree $\chi_L$-polynomial  in the denominator of
Eq. \eqref{SD-explicit}. The coefficients $c_\pm$ can then be found from Eqs. \eqref{c-coeffs}.
Expanding in $1/N$ up to NTLO, after simple algebra we arrive at the following results
\begin{subequations}\label{massless-coeffs-massses}
\begin{eqnarray}
 c_+ &=& 1 - \frac{5}{16N},\\
 c_- &=& \frac{5}{16N},\\
 m_+^2 &=& m_0^2 \left( 1 + \frac{1}{4N} \right), \\
 m_-^2 &=& 5 m_0^2 \left( 1 +   \frac{1}{4N} \right).
\end{eqnarray} 
\end{subequations}
Going back to dS we obtain
the final result for the two-point correlator at this order, that can be written as the sum of two free, massive propagators,
\begin{equation}
 G(x,x') = c_+ \, \Delta^{(m_+)}(x,x') + c_- \, \Delta^{(m_-)}(x,x').
\end{equation}
In the massless case the coefficients and masses are given by Eqs.~\eqref{massless-coeffs-massses}. This coincides with a previous result obtained in Ref.\cite{Serreau2}
using a low-momentum expansion in dS.

All the way through this calculation we have neglected terms that decay faster than the leading ones in the IR, essentially by appealing to the known behaviour for the free propagator $\Delta(x,x')$. When looking at the result for the full propagator $G(x,x')$ we have obtained at NTLO in $1/N$, one would be tempted to drop the term involving the mass $m_-$ in the IR limit in the same vein, as it would decay faster that the one with $m_+$. Notice however  that  the size of the ``mass'' that controls the decay of the terms we have dropped in $\Delta(x,x')$ and subsequent expressions is given by $H$, while  both $m_+$ and $m_-$  are of order $ {\lambda}^{1/4} H$. In  other words, the ratio between  $m_+$ and $m_-$ is fixed independently of $\lambda$ and therefore both masses are parametrically smaller than $H$. 

The importance of keeping the term with $m_-$ becomes evident when computing $m_{dyn}^2$ by evaluating Eq.~\eqref{SD-2props} at $L=0$. Both terms are needed already at the LO in $\sqrt{\lambda}$ to be able to reproduce Eq.~\eqref{mdyn}, while clearly any term with the form of a free propagator with a squared mass of order $H^2$ would only give higher order corrections. This same situation arises when going to higher orders in $1/N$. Indeed, one of the contributions to the self-energy $\Sigma$ at NNTLO will be given by the already included diagrams evaluated using $G^{(NTLO)}$ instead of $G^{(LO)}$, where mantaining both terms with $m_+$ and $m_-$ will be necessary. Of course, there will also be contributions coming from new kinds of diagrams.

In conclusion, the SDE on the sphere can be solved at NTLO in $1/N$ by properly approximating the self-energy in the IR.  The IR limit corresponds to the
structure of the momentum coefficients $G_L$ in the complex $L$-plane.

\section{The four-point correlation functions}

Here we apply some of the results obtained so far in this paper to analyze the four-point correlation functions  in the IR.
In dS spacetime, the four-point correlation functions can be written in terms of the four-point vertex functions, $\Gamma_{abcd}^{(4)}( \{ x_i' \})$, as
%  \begin{equation}\label{4point} 
% G_{abcd}^{(4)}( \{ x_i \})=  \int_{ \{ x_i' \}}G_{aa'}(x_1,x_1')G_{bb'}(x_2,x_2')G_{cc'}(x_3,x_3')G_{dd'}(x_4,x_4')\Gamma_{a'b'c'd'}^{(4)}( \{ x_i' \})
% \end{equation} 
\begin{equation}\label{4point} 
G_{abcd}^{(4)}( \{ x_i \})=  \int_{ \{ x_i' \}}G(x_1,x_1')G(x_2,x_2')G(x_3,x_3')G(x_4,x_4')\Gamma_{abcd}^{(4)}( \{ x_i' \})
\end{equation}
where $( \{ x_i \})$ stands for $(x_1,x_2,x_3,x_4)$.

As emphasized in \cite{Serreau13} in dS, the  four-point vertex functions in the large N limit can be written in terms of the two-point function of the composite field $\chi=\frac{\lambda\phi_a\phi_b}{2N}$, namely $D(x,x')$, as 
%\cite{foot2} 
 \begin{equation}
\Gamma_{abcd}^{(4)}( \{ x_i \})=\delta_{ab}\delta_{cd}\delta^{(d)}(x_1,x_2) \delta^{(d)}(x_3,x_4) D(x_1,x_3)+ c_{\rm{perm}},
\end{equation} where ``$ c_{\rm{perm}}$'' stands for the two cyclic permutations needed to make $\Gamma_{abcd}^{(4)}$ symmetric (recall that our coupling constant $\lambda$ should be divided by 3 to match the one used in \cite{Serreau13}). The two-point function $D(x_1,x_2)$ is given by a local contribution (which corresponds to the tree-level contribution, but with full propagators) plus a nonlocal part that involves the resummed bubble-chain $I(x,x')$, 
%we have already computed:
% , that contribute to the two-point functions of the original fields $\phi_a$:  
\begin{equation}
 D(x,x')=-\frac{\lambda}{N}[\delta^{(d)}(x,x')+ I(x,x')].
\end{equation}
Inserting this into Eq.\eqref{4point} we decompose the four-point function as
 \begin{equation}
G_{abcd}^{(4)}( \{ x_i \})= G_{abcd}^{(4,\rm{tree})}( \{ x_i \}) + 
 G_{abcd}^{(4,\rm{loop})}( \{ x_i \})\, .
 \end{equation}
% where $I(x,x')$ is the resummed bubble-chain,
% \begin{equation}
%   I(x,x')= \Pi(x,x')+i\int_z \Pi(x,z) I(z,x'),
%   \end{equation}with  $\Pi(x,x')$ the single bubble diagram. 
  
Our goal is to analyze the  four-point  functions (\ref{4point}) in the IR limit, meaning the case when the four points are far apart one from the other, and at leading order in $1/N$.  Recall that at leading order in $1/N$, the propagator   $G(x,x')$ is given by a free massive propagator with mass $m_0$, which we are writing as $ \Delta^{(m_0)}(x,x')$. 
 Therefore, to understand the IR behaviour we can use the results obtained in \cite{M&M, Hollands2} for the case of interacting massive fields. There it is shown that for generic $n$-point correlators involving loops, when two or several  points are far apart from each other, the  correlator decays at least as  fast as $r^{-\frac{m_0^2}{(d-1) H^2}+\mathcal{O}(\epsilon)}$, where $\epsilon$ is an infinitesimal positive constant and $r$ is the maximum distance between a pair of points.
  
The resummed bubble-chain $I(x,x')$ decays, in the IR limit,  as a massive propagator with mass $2 m_0$ (see Eq.\eqref{IeucRes}).
This suggests that the loop corrections might decay  faster than the tree level contribution.  Indeed,  the fully Lorentzian calculation performed in \cite{Serreau4poit}  shows that there is a scaling behaviour in the loop contributions to the  four-point functions that is different from the tree level part. To show this explicitly,   we work  in conformal coordinates ($ds^2=(H\eta)^{-2}[-d\eta^2+\delta_{ij} dx^i dx^j]$) with $d=4$, and compute the Fourier transform of their results in momentum space to compute the  corresponding correlators  at equal times. Using that

  \begin{equation}
G_{abcd}^{(4)}( \eta,\{  \vec{k}_i \}) (2\pi)^3 (-H \eta)^{-3}\delta\left({\sum_i \vec{k}_i}\right)=\int_{ \{ \vec{x}_i \} }\,\exp\{i\sum_i  \vec{k^i}\cdot \vec{x^i}\} G_{abcd}^{(4)}( \eta, \{ \vec{x}_i \}),
  \end{equation} where $\int_{\vec{x}}\equiv\int (-H \eta)^{-3}  d^3x$, and  
  \begin{equation}
\frac{ (-H \eta)^{-3}}{|\vec{x}|^w} =\frac{\Gamma\left[\frac{3-w}{2}\right]}{2^w\pi^{\frac{3}{2}}\Gamma\left[\frac{w}{2}\right]}\int_{\vec{k}} \,\exp\{i  \vec{k }\cdot \vec{x}\}  |\vec{k}|^{w-3},
  \end{equation} it is immediate to see that the tree level contribution behaves as

  \begin{eqnarray}\label{4pointapptree} 
G_{abcd}^{(4,\rm{tree})}( \{ x_i \})&\sim&     \delta_{ab} \delta_{cd}\frac{\lambda  (\eta^{2})^{\frac{3m_{0}^2}{(d-1)H^2}} }{ N m_{0}^8V_d^3}  \left[x_{21}^{-\frac{2m_{0}^2}{(d-1)H^2}}x_{31}^{-\frac{2m_{0}^2}{(d-1)H^2}}x_{41}^{-\frac{2m_{0}^2}{(d-1)H^2}}+ \rm{permutations}\right], 
\end{eqnarray} decaying as a product of free propagators with mass $m_0$. On the other hand, the loop contribution has two pieces: one scales as  the tree level one,
while the other behaves as the product of two propagators of mass $m_0$ and one of mass $2m_0$,

  \begin{eqnarray}\label{4pointapp} 
G_{abcd}^{(4,\rm{loop})}( \{ x_i \})&\sim&     \delta_{ab} \delta_{cd} \frac{\lambda (\eta^{2})^{\frac{3m_{0}^2}{(d-1)H^2}}}{N m_{0}^8V_d^3}  \left[x_{21}^{-\frac{2m_{0}^2}{(d-1)H^2}}x_{43}^{-\frac{2m_{0}^2}{(d-1)H^2}}x_{13}^{-\frac{8m_{0}^2}{(d-1)H^2}}+  {\rm{permutations}}\right]. 
\end{eqnarray}  
This is related to the IR behavior of $I(x,x')$.
  
\section{Conclusions}

In this paper we presented a novel approach to compute the NTLO corrections to the two-point functions of quantum fields in dS spacetime, in the deep IR limit.
Our approach is based on the fact that the IR behavior of the two-point functions in dS spacetime is related with the poles of the analytically continued two-point functions
on the sphere. The SDE that determines the two-point functions on the sphere involves the self-energy, that can be  computed 
knowing the product of  two massive propagators. In the deep IR, this product  can be approximated by a single propagator with a mass given by the sum of the masses.
Going to the sphere, solving there the SDE, and coming back to dS we have been able to obtain the NTLO corrections in $1/N$ to the two-point functions.
As pointed out earlier in Ref.~\cite{Serreau2}, this corresponds to a linear combination of massive propagators with masses given by $m_\pm$. 
In the present approach, it is clear that the fact that the corrected propagator is a linear combination
of propagators with different masses comes from the IR behavior of the self-energy, which decays with a squared mass $5 m_0^2$. 

It is worth noticing that the $1/N$ expansion provides a reorganization of perturbation theory, by collecting an infinite subset of diagrams which individually grow secularly at large distances. This can be explicitly seen by comparing the different   bubble diagrams in Fig.~\ref{fig:1}. For example the single   bubble diagram is given by Eq. \eqref{single-bubble}, which in the IR can be approximated using Eq. \eqref{IR-DeltaDelta}, giving
\begin{equation} \label{single-bubble-explicit}
 \Pi(x, x') \simeq -2 m_0^2 \, \Delta^{(\sqrt{2}m_0)}(x, x').
\end{equation}
Using this result, the two-bubble diagram instead goes as
\begin{equation} \label{two-bubble-explicit}
 \int_z \Pi(x, z) \Pi(z, x') \simeq 4 m_0^4 \, \int_z \Delta^{(\sqrt{2}m_0)}(x, z) \Delta^{(\sqrt{2}m_0)}(z, x') = - (2m_0^2)^2 \, \frac{\partial \Delta^{(m)}(x, x')}{\partial m^2} \Bigg|_{m = \sqrt{2}m_0},
\end{equation} and  the one  of the n-bubble diagram as 
\begin{equation} \label{n-bubble-explicit}
 \int_{z_1,\dots,z_{n-1}} \Pi(x, z_1) \dots \Pi(z_{n-1}, x') \simeq  -\frac{ (2m_0^2)^n}{n!} \, \frac{\partial ^{n-1}\Delta^{(m)}(x, x')}{\partial (m^2)^{n-1}} \Bigg|_{m = \sqrt{2}m_0}.
\end{equation}  Here  the  last equalities follow from a  generic property of the massive propagator (see Ref. \cite{nos}).
When explicitly computing the $m^2$-derivative of the free massive propagator using Eq. \eqref{IRfree}, it is easy to see that at large $r$, the n-bubble diagram \eqref{n-bubble-explicit}  grows with a  factor  $(\log r)^{n-1}$ with respect to the single-bubble one \eqref{single-bubble-explicit}.  

In our approach, we were able to perform the resummation of all these secularly growing corrections at NTLO, obtaining a correction that remains subleading at large separations, Eq. \eqref{IeucRes}. The IR behaviour cannot be made evident on a compact space, before analytic continuation.  The crucial point is the observation that the IR limit of the Lorentzian propagators is determined by the pole structure of the Euclidean propagators around $L=0$ in the complex $L$-plane. This property simplifies drastically the study of the large distance behaviour of the analitycally continued propagators. It would we worth to generalize to higher order in the $1/N$ expansion. Iterating the procedure, we expect additional poles at the NNTLO coming from the IR behavior of the self-energy computed with the corrected propagators.

As a straightforward application of our results, we have discussed  the IR behaviour of the (equal-time) four-point correlation functions. We used the fact that, in the large $N$
limit, the four-point correlators can be written in terms of integrals of free massive propagators.  
This allowed us to use previously obtained IR bounds for $n$-point correlators for massive fields \cite{M&M,Hollands2}.

\acknowledgments
% \section{Acknowledgements}
This work has been supported by CONICET,  ANPCyT, UBA and UNCuyo.

\end{document}